\documentclass{article}

\usepackage[a4paper,margin=3.0cm]{geometry} 

\title{sumo3Dviz: A three dimensional traffic visualisation}
\author{
    {\small
        Kevin Riehl \orcidlink{0000-0003-4620-8379}
        \thanks{\small Traffic Engineering Group, Institute for Transport Planning and Systems, ETH Zurich, Stefano-Franscini-Platz 5, 8093 Zurich, Switzerland. Correspondence: kriehl@ethz.ch}
        \quad
        Julius Schlapbach \orcidlink{0009-0006-0118-6451}
        \quad
        Anastasios Kouvelas \orcidlink{0000-0003-4571-2530}
        \quad
        Michail A. Makridis \orcidlink{0000-0001-7462-4674}
    }
}

\date{}  

\usepackage{xfrac}
\usepackage{orcidlink} 
\usepackage{booktabs}
\usepackage{pifont}
\usepackage{placeins}
\usepackage{url}
\usepackage{hyperref}
\usepackage[backend=biber]{biblatex}
\begin{document}

\maketitle

\begin{abstract}
Traffic microsimulation software such as SUMO generate rich spatio-temporal data describing individual vehicle movements, interactions, and support the development of control strategies. 
While numerical outputs and 2D visualisations are sufficient for many technical analyses, they are often inadequate for applications that require intuitive interpretation, effective communication, or human-centred evaluation. 
In particular, user studies in mobility psychology, acceptance research, and virtual experience stated-preference experiments require realistic visualisations that reflect how traffic scenarios are perceived from a human perspective.
This paper introduces sumo3Dviz, a lightweight, open-source 3D visualisation pipeline for SUMO traffic simulations. It converts standard SUMO simulation outputs, such as vehicle trajectories and signal states, into high-quality 3D renderings using a Python-based framework. 
In contrast to heavyweight game-engine-based approaches or tightly coupled co-simulation frameworks, sumo3Dviz is designed to be simple, scriptable, and reproducible. 
The tool is installable through the pip package manager, runs across operating systems, and works independently of any proprietary software or licenses.
sumo3Dviz supports both external camera views and first-person perspectives, enabling cinematic overviews as well as driver-level experiences. 
The rendering process is optimized for batch video generation, making it suitable for large-scale scenario visualisation, educational demonstrations, and automated experiment pipelines. 
A key technical challenge addressed by the tool is trajectory interpolation and orientation smoothing, enabling visually coherent motion from discrete simulation outputs. \\
Source Code on project's GitHub page:  \url{https://github.com/DerKevinRiehl/sumo3dviz/}. \\
\textbf{Keywords:} 3D Visualisation, Mobility Psychology, User Experience, Virtual Experience Stated Preference
\end{abstract}

\section{Introduction}

Traffic microsimulation has become essential in transportation research, planning, and operations~\cite{shahdah2015application,hollander2008principles}.
Simulators such as SUMO~\cite{krajzewicz2012recent} enable the detailed modelling of individual vehicles, infrastructure, and control strategies at high spatial and temporal resolution.
These simulations generate large volumes of data describing trajectories, speeds, interactions, and signal states.
While suited for quantitative analysis, these outputs often lack intuitive interpretability~\cite{mahmassani201650th,mahmud2019micro}.

Visualisation bridges this gap by making microsimulation results accessible and meaningful~\cite{hughes2004visualization,klein2009visualization}.
In particular, visual representations support:
\begin{itemize}
    \item enhancing interpretability of simulation outcomes by transforming abstract values into observable traffic scenes,
    \item debugging and model validation through the detection of unrealistic behaviour or configuration errors,
    \item communicating research results to non-technical stakeholders and decision-makers,
    \item teaching and dissemination in academic and applied contexts.
\end{itemize}

For many applications, two-dimensional visualisations are insufficient~\cite{erath2017visualizing,juric2024real}.
Human-centred evaluation, user studies, and virtual experience stated preference experiments require stimuli that approximate real-world perception~\cite{facchini2025virtual}.
Three-dimensional visualisation enables street-level context, depth cues, and ego-centric viewpoints that better reflect how traffic situations are experienced by users~\cite{weber2024visualanalytics}.
These perspectives are particularly relevant when studying comfort, trust, acceptance, or perceived safety of traffic systems.
Beyond user studies, 3D traffic visualisation is also a key component of emerging digital twin concepts for urban mobility~\cite{batty2018digital}, where it acts as an interface between complex simulation models and human decision-making.

Existing 3D visualisation approaches for traffic microsimulation often rely on complex game engines or tightly coupled co-simulation frameworks.
These solutions typically require high setup effort, raise reproducibility concerns, suffer from compatibility issues across operating systems, and depend on proprietary software.
Such characteristics hinder their use in automated workflows, teaching, and experimental research.

This work presents \texttt{sumo3Dviz}, a lightweight, open-source three-dimensional visualisation pipeline for SUMO traffic simulations.
The tool converts standard SUMO output log files into high-quality 3D renderings using a Python-based framework.
It is designed to be simple, scriptable, and platform-independent, while supporting both external camera views and ego-centric perspectives.
The rendering process is optimised for batch video generation, enabling its integration into large-scale simulation and experiment pipelines.
The tool supports all major operating systems, including Windows, macOS, and various Linux distributions, and installs via pip in standard Python-environments.

The remainder of this paper is organised as follows.
Section~2 reviews related work on three-dimensional traffic visualisation and systematically compares existing tools.
Section~3 describes the software architecture and design principles of \texttt{sumo3Dviz}.
Section~4 presents methodological details on the 3D scene graph and vehicle trajectory smoothing.
Section~5 demonstrates an example use case and discusses performance aspects.
Finally, Section~6 concludes the paper and outlines future work.

\section{Related Work}

A variety of tools and pipelines exist for visualising traffic microsimulations, ranging from integrated simulators to external game-engine-based solutions.
These approaches differ substantially in terms of realism, usability, platform-dependence, reproducibility, technical feasibility, and integration effort, as shown in Table~\ref{tab:viz_comparison}.

\begin{table*}[t]
    \centering
    \caption{Comparison of existing traffic microsimulation visualisation tools.}
    \label{tab:viz_comparison}
    \renewcommand{\arraystretch}{1.2}
    \scriptsize
    \begin{tabular}{lcccccccc}
    \toprule
    \textbf{Tool} &
    \textbf{3D} &
    \textbf{Ego View} &
    \textbf{Open Source} &
    \textbf{Lightweight} &
    \textbf{SUMO-native} \\
    \midrule
    
    \href{https://eclipse.dev/sumo/}{SUMO-GUI}~\cite{krajzewicz2012recent} &
    \ding{51} &
    -- &
    \ding{51} &
    \ding{51} &
    \ding{51} \\
    
    \href{https://github.com/patmalcolm91/SumoNetVis}{SumoNetVis} &
    -- &
    -- &
    \ding{51} &
    \ding{51} &
    \ding{51} \\
    
    \href{https://www.dspace.com/de/gmb/home/products/sw/experimentandvisualization/aurelion_sensor-realistic_sim.cfm}{AURELION} &
    \ding{51} &
    -- &
    -- &
    -- &
    \ding{51} \\
    
    \href{https://github.com/SvenMertin/SUMO3d}{SUMO3D} &
    \ding{51} &
    -- &
    \ding{51} &
    -- &
    \ding{51} \\
    
    \href{https://traffic3d.org/index.html}{Traffic3D} &
    \ding{51} &
    -- &
    \ding{51} &
    -- &
    \ding{51} \\
    
    \href{https://github.com/TUM-VT/Sumonity}{Sumonity}~\cite{pechinger2024sumonity} &
    \ding{51} &
    -- &
    \ding{51} &
    -- &
    \ding{51} \\
    
    \href{https://github.com/SimuTraffX-Lab/SUMO2Unity}{Sumo2Unity}~\cite{mohammadi2024sumo2unity,mohammadi2025open} &
    \ding{51} &
    \ding{51} &
    \ding{51} &
    -- &
    \ding{51} \\
    
    \href{https://github.com/AugmentedDesignLab/Sumo2Unreal}{Sumo2Unreal}~\cite{paranjape2020modular,youssef2024joining} &
    \ding{51} &
    \ding{51} &
    \ding{51} &
    -- &
    \ding{51} \\
    
    \href{https://carla.org/}{CARLA + SUMO}-based~\cite{dosovitskiy2017carla} &
    \ding{51} &
    \ding{51} &
    \ding{51} &
    -- &
    -- \\
    
    \href{https://www.ptvgroup.com/de/produkte/ptv-vissim}{VISSIM 3D}~\cite{fellendorf2010microscopic} &
    \ding{51} &
    \ding{51} &
    -- &
    -- &
    -- \\
    
    \href{https://www.aimsun.com/}{AIMSUN 3D} &
    \ding{51} &
    -- &
    -- &
    -- &
    -- \\
    
    \midrule
    \textbf{\href{https://github.com/DerKevinRiehl/sumo3Dviz}{sumo3Dviz} (this work)} &
    \ding{51} &
    \ding{51} &
    \ding{51} &
    \ding{51} &
    \ding{51} \\
    \bottomrule
    \end{tabular}
\end{table*}

\subsection{Built-in and Lightweight Visualisation Tools}

\href{https://eclipse.dev/sumo/}{SUMO}~\cite{krajzewicz2012recent} provides a built-in graphical user interface that supports two-dimensional visualisation of networks, vehicles, and signal states.
Being tightly integrated with the simulation engine, the SUMO-GUI is lightweight and responsive.
It is well suited for debugging, model validation, and quick inspection of simulation behaviour.
However, it is limited to 2D representations, offers restricted camera control, and does not support street-level or ego-centric perspectives.
The visual realism in the preliminary 3D rendering is minimal, making it unsuitable for human-centred user studies or cinematic video generation.

Lightweight tools such as \href{https://github.com/patmalcolm91/SumoNetVis}{SumoNetVis} provide simple, scriptable two-dimensional visualisations of SUMO networks and trajectories.
They are typically pip-installable and easy to integrate into Python workflows.
While these tools improve accessibility and reproducibility, they remain limited to abstract 2D representations and are not intended for immersive or perceptual evaluation.

\subsection{Dedicated 3D Renderers for SUMO}

Several projects aim to provide three-dimensional visualisation directly from SUMO simulations. For example, 
\href{https://www.dspace.com/de/gmb/home/products/sw/experimentandvisualization/aurelion_sensor-realistic_sim.cfm}{AURELION} offers a full 3D rendering environment with support for textures, lighting, and external camera views, providing a more realistic representation of traffic scenes when compared to SUMO-GUI.
However, it requires substantial setup effort, relies on specialised environments, and lacks Python scripting interfaces for automated workflows.

Other academic or community-driven projects, such as \href{https://github.com/SvenMertin/SUMO3d}{SUMO3D} or \href{https://traffic3d.org/index.html}{Traffic3D}, demonstrate three-dimensional traffic rendering with varying degrees of realism.
These tools typically serve as research prototypes or focus on specific demonstration scenarios.
Installation complexity, limited documentation, and lack of maintained scripting interfaces hinder a broad adoption.

\subsection{Game Engine-Based Pipelines}

A common approach to high-quality traffic visualisation is the use of game engines such as Unity or Unreal Engine.
Several pipelines convert SUMO trajectories to formats compatible with these environments, including \href{https://github.com/TUM-VT/Sumonity}{Sumonity}~\cite{pechinger2024sumonity}, \href{https://github.com/SimuTraffX-Lab/SUMO2Unity}{Sumo2Unity}~\cite{mohammadi2024sumo2unity,mohammadi2025open}, and \href{https://github.com/AugmentedDesignLab/Sumo2Unreal}{Sumo2Unreal}~\cite{paranjape2020modular,youssef2024joining}.
These approaches provide realistic environments, advanced lighting, and virtual reality support.
They have been applied to digital twin projects~\cite{amini2023integrating} and immersive demonstrations.
Despite their visual quality, game-engine-based pipelines typically involve substantial engineering effort.
They require proprietary software, involve complex data conversion, and require manual scene configuration, making reproducibility and portability challenging.

\href{https://carla.org/}{CARLA}-based~\cite{dosovitskiy2017carla} co-simulation frameworks combine SUMO with photorealistic 3D environments and advanced sensor models.
While highly suitable for autonomous driving research, such frameworks are computationally resource-intensive and exceed the requirements of most traffic visualisation tasks.
They are not designed for lightweight rendering or high-throughput video generation workflows.

\subsection{High-End and Proprietary Simulation Platforms}

Commercial tools such as \href{https://www.ptvgroup.com/de/produkte/ptv-vissim}{VISSIM}~\cite{fellendorf2010microscopic} and \href{https://www.aimsun.com/}{AIMSUN}~\cite{casas2010traffic} provide integrated three-dimensional visualisation capabilities.
These platforms offer mature graphical interfaces and industry-standard workflows.
However, they depend on proprietary licenses and closed ecosystems, which limits their integration into automated, reproducible research workflows and open-source toolchains.

\subsection{Positioning and Contribution}

The review above highlights a gap between lightweight but abstract visualisation tools and highly realistic but resource-intensive simulation environments.
To the best of our knowledge, there is currently no open-source, platform-independent, pip-installable Python tool that directly consumes SUMO simulation outputs and produces lightweight, three-dimensional visualisations optimised for scripting and batch video generation.

\texttt{sumo3Dviz} addresses this gap by providing:
\begin{itemize}
    \item direct processing of standard SUMO trajectory and signal outputs,
    \item fast three-dimensional rendering without reliance on game engines,
    \item support for both external camera views and ego-centric perspectives,
    \item basic environmental context including buildings, vegetation, and signals,
    \item trajectory interpolation and orientation smoothing for visually coherent motion,
    \item a lightweight, reproducible, and fully scriptable Python-based workflow.
\end{itemize}

By prioritising simplicity, automation, platform-independence, and perceptual relevance, \texttt{sumo3Dviz} complements existing tools and enables new applications in teaching, communication, and human-centred traffic research.

\section{Software Architecture}

\begin{figure}[!ht]
    \centering
    \caption{Overview of the \texttt{sumo3Dviz} workflow, from SUMO inputs to rendered video output.}
    \label{fig:workflow}
    \includegraphics[width=1.0\linewidth]{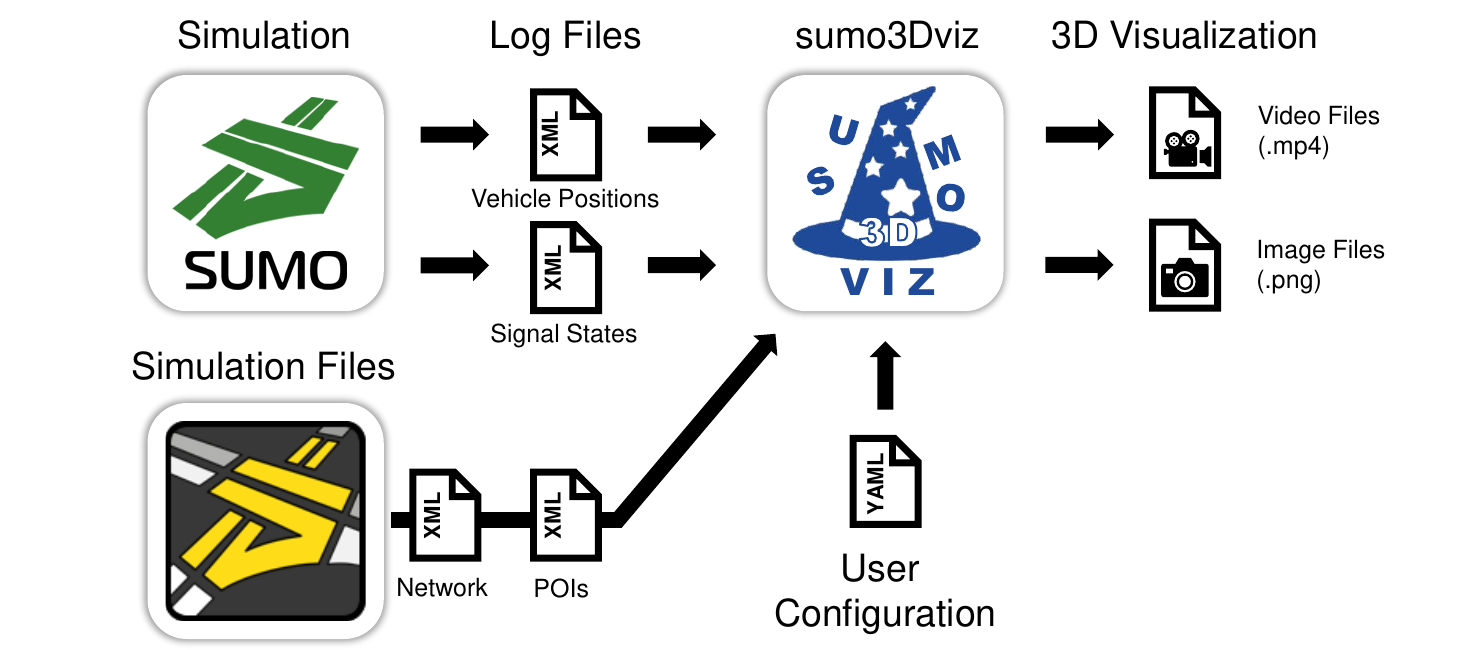}
\end{figure}

\texttt{sumo3Dviz} is designed as a lightweight and self-contained visualisation pipeline that operates directly on standard SUMO simulation inputs and outputs.
The tool requires no modification of the simulation engine and does not rely on co-simulation or external rendering frameworks.
Its primary output consists of rendered videos representing three-dimensional traffic scenes. The basic workflow is illustrated in Figure~\ref{fig:workflow}.

\subsection{Input Data and Preprocessing}

The input to \texttt{sumo3Dviz} consists of SUMO network and configuration files, as well as the simulation output logs.
The network input includes the road network definition and optional polygon files.
Polygon files specify static scene elements such as buildings, fences, trees, or other points of interest.

The simulation output logs consist of a vehicle trajectory file containing time-stamped positions and headings.
Optionally, signal state logs may be provided to visualise traffic light behaviour.
These outputs can be generated by SUMO with minor adjustments to the simulation configuration and do not require additional post-processing or instrumentation.

During preprocessing, discrete vehicle trajectories are parsed and transformed into continuous motion representations.
Trajectory interpolation and orientation smoothing are applied to ensure visually coherent movement despite the discrete simulation time steps.
Static elements are loaded once and remain fixed throughout the rendering process.

\subsection{Configuration and Usage}

Visualisation behaviour is controlled through user-defined configuration files in YAML format.
These files specify input paths, rendering modes, camera parameters, visual styles, and output settings, enabling reproducible rendering and simplifying the integration into automated workflows.

\texttt{sumo3Dviz} can be installed in a Python virtual environment using \texttt{pip}.
Basic usage is provided through a command-line interface, allowing users to generate visualisations without programming experience.
For advanced applications, the library can be imported as a Python module and extended programmatically.

\subsection{Rendering Modes}

The tool supports four primary rendering modes.
\begin{itemize}
    \item In the \textit{Eulerian mode}, scenes are rendered from a fixed external camera perspective.
    \item In the \textit{Lagrangian mode}, the camera is attached to a specific vehicle, enabling ego-centric views.
    \item In the \textit{Cinematic mode}, the camera follows a predefined three-dimensional trajectory, supporting fly-throughs and controlled viewpoints.
    \item In the \textit{Interactive mode}, users freely navigate static scenes using keyboard controls, with dynamic objects disabled.
\end{itemize}
Each mode supports rendering of single images or time-resolved video sequences.
The rendering pipeline is optimised for batch processing and headless execution.

\subsection{Visual Customisation}

The visual appearance of scenes is highly configurable.
Users can specify textures for sky and ground surfaces, colour schemes and width constraints for road segments, and set appearance parameters for static objects.
Multiple open-source and royalty-free vehicle models are supported and can be assigned dynamically.
This flexibility enables visual styles ranging from abstract representations to more realistic street-level scenes, depending on the application context.

\section{Methods}

\subsection{3D Scene Graph}

The proposed visualisation software renders a three-dimensional visualisation, where the scene graph consists of following, customisable components:
\begin{itemize}
    \item camera
    \item light source
    \item sky (textured sphere)
    \item ground (textured floor)
    \item road network (rendered from SUMO network file)
    \item static objects (trees, freeway fences, traffic lights, shops, homes, building blocks) 
    \item moving objects (ego vehicle, other vehicles)
\end{itemize}

The camera's position is customisable in four different perspectives: 
(i) fixed position (Eulerian mode), 
(ii) attached to a specific vehicle (ego view, Lagrangian mode), 
(iii) following a predefined three-dimensional trajectory (Cinematic mode),
(iv) interactive control (user can navigate static scenes using keyboard controls).

\begin{figure}[!ht]
    \centering
    \includegraphics[width=1.0\linewidth]{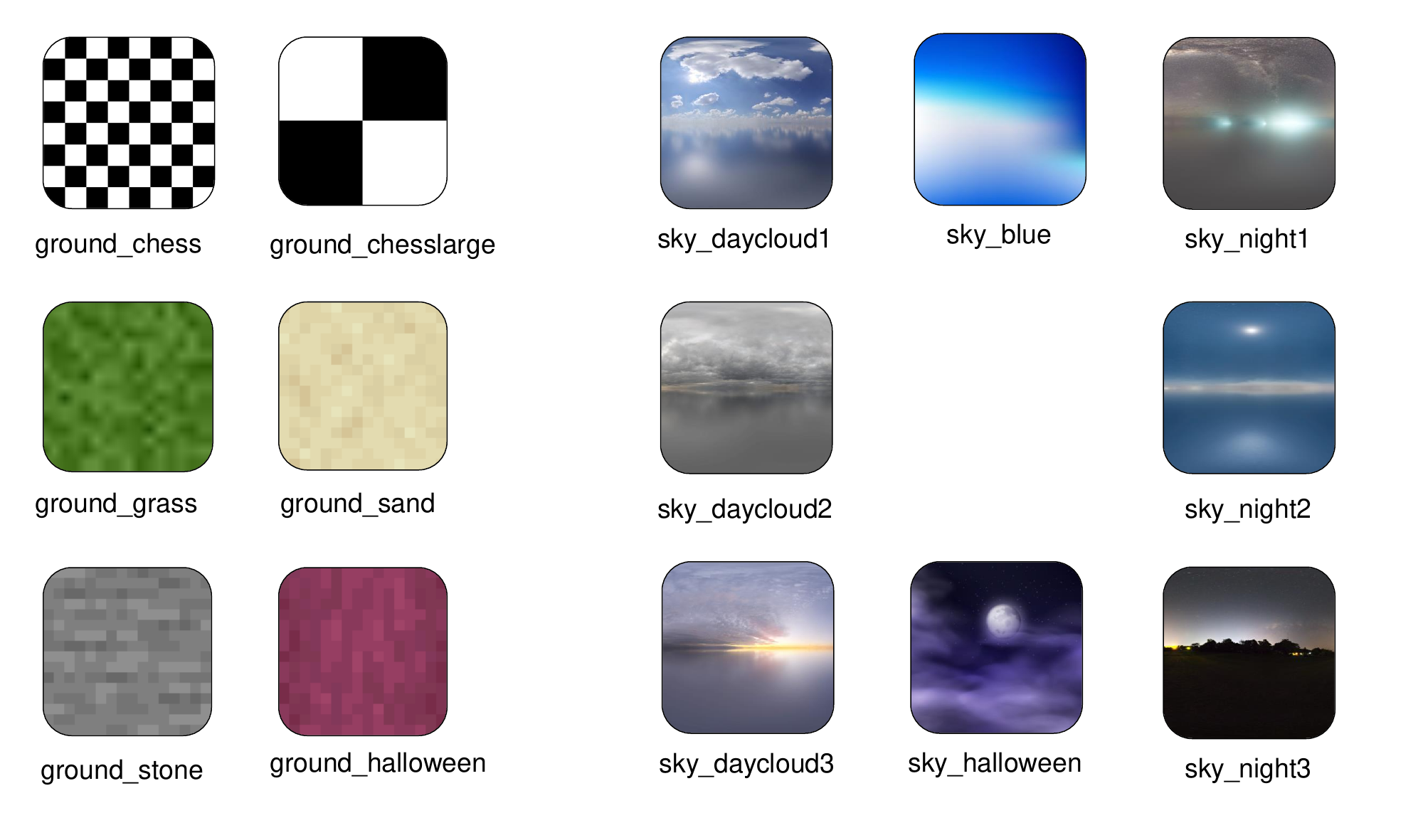}
    \caption{Available textures for ground and sky rendering in \texttt{sumo3Dviz}.}
    \label{fig:textures}
\end{figure}

The sky is a sphere, with a customisable texture, selectable from a provided library (see Figure~\ref{fig:textures}), including:
\textit{`sky\_blue`} (clear blue sky),
\textit{`sky\_daycloud1`}, 
\textit{`sky\_daycloud2`}, 
\textit{`sky\_daycloud3`} (cloudy sky),  
\textit{`sky\_night1`}, 
\textit{`sky\_night2`},
\textit{`sky\_night3`} (night sky with stars),
\textit{`sky\_halloween`} (halloween-themed sky).

The ground is an area, with a customisable texture, selectable from a provided library (see Figure~\ref{fig:textures}), including:
\textit{`ground\_grass`} (natural grass texture),
\textit{`ground\_stone`} (stone/concrete texture),
\textit{`ground\_sand`} (sand/desert texture),
\textit{`ground\_chess`} (chess board pattern, small),
\textit{`ground\_chesslarge`} (chess board pattern, large),
\textit{`ground\_halloween`} (halloween-themed ground).

\FloatBarrier

\begin{figure}[!ht]
    \centering
    \includegraphics[width=1.0\linewidth]{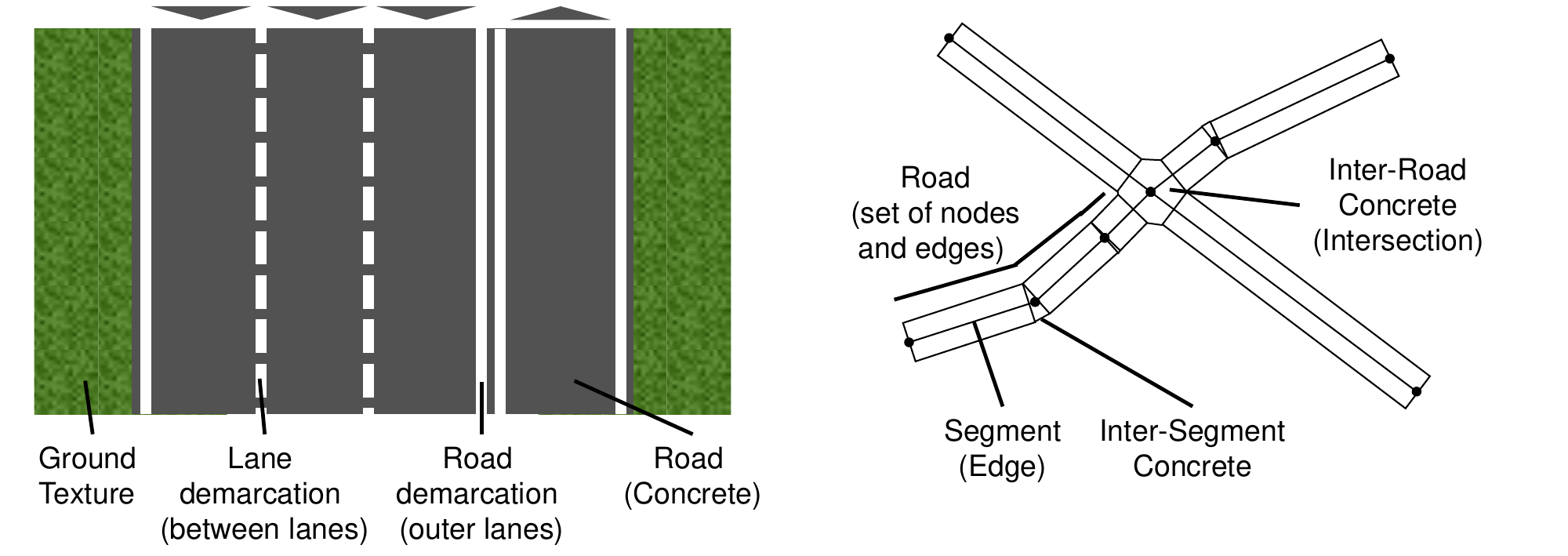}
    \caption{Road network rendering approach: layered structure (left) and geometric construction from SUMO network (right).}
    \label{fig:road_rendering}
\end{figure}

The road rendering consists of three components, as shown on the left in Figure~\ref{fig:road_rendering}.
The ground area layer (e.g. grass) is overlaid with a road polygon (e.g. gray concrete), on top of which road and lane demarcation lines (e.g. white separators) are rendered.
The road network is rendered based on a SUMO network file, as shown on the right in Figure~\ref{fig:road_rendering}.
Each road (potentially with multiple lanes) in the SUMO network graph comprises multiple edges defining its geometric path.
Each edge segment on that path (including all lanes) is rendered as a rectangle, with triangular polygons connecting adjacent segments.
At intersections, the geometric polygon from SUMO intersections (inter-road concrete) is used for rendering.

\FloatBarrier

\begin{figure}[!ht]
    \centering
    \includegraphics[width=1.0\linewidth]{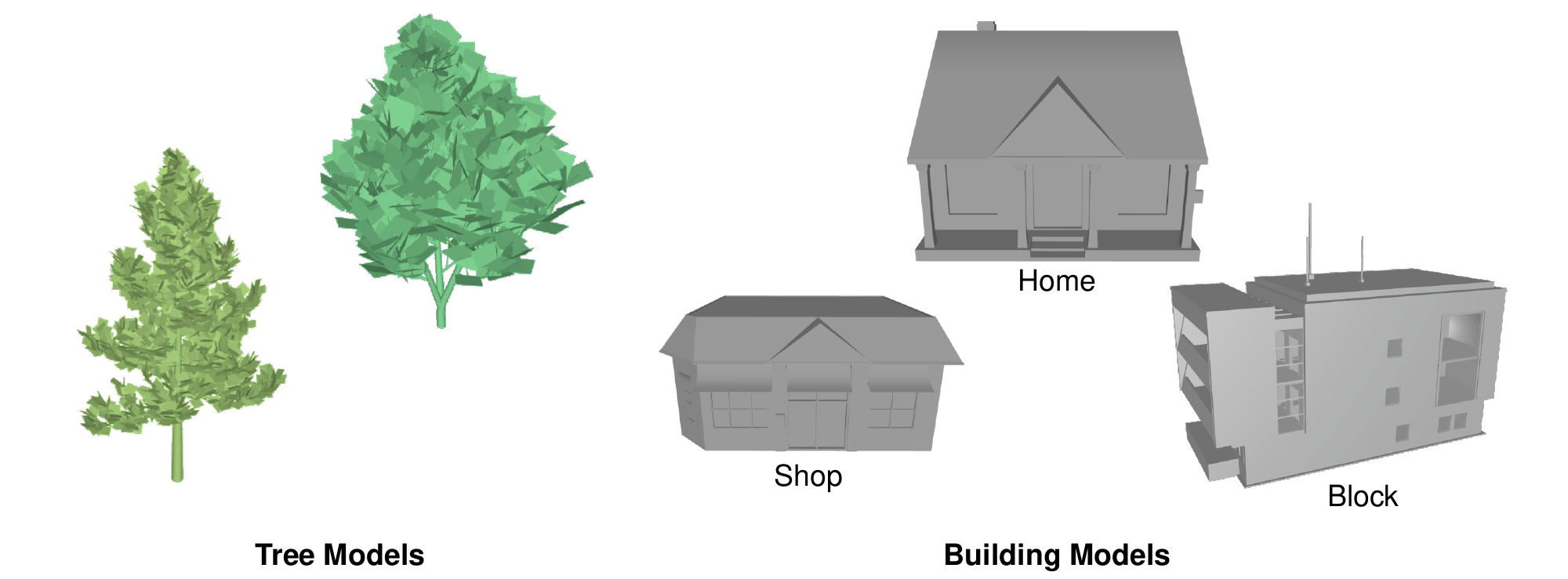}
    \caption{Static objects in \texttt{sumo3Dviz} (trees and buildings).}
    \label{fig:static_objects}
\end{figure}

\begin{figure}[!ht]
    \centering
    \includegraphics[width=1.0\linewidth]{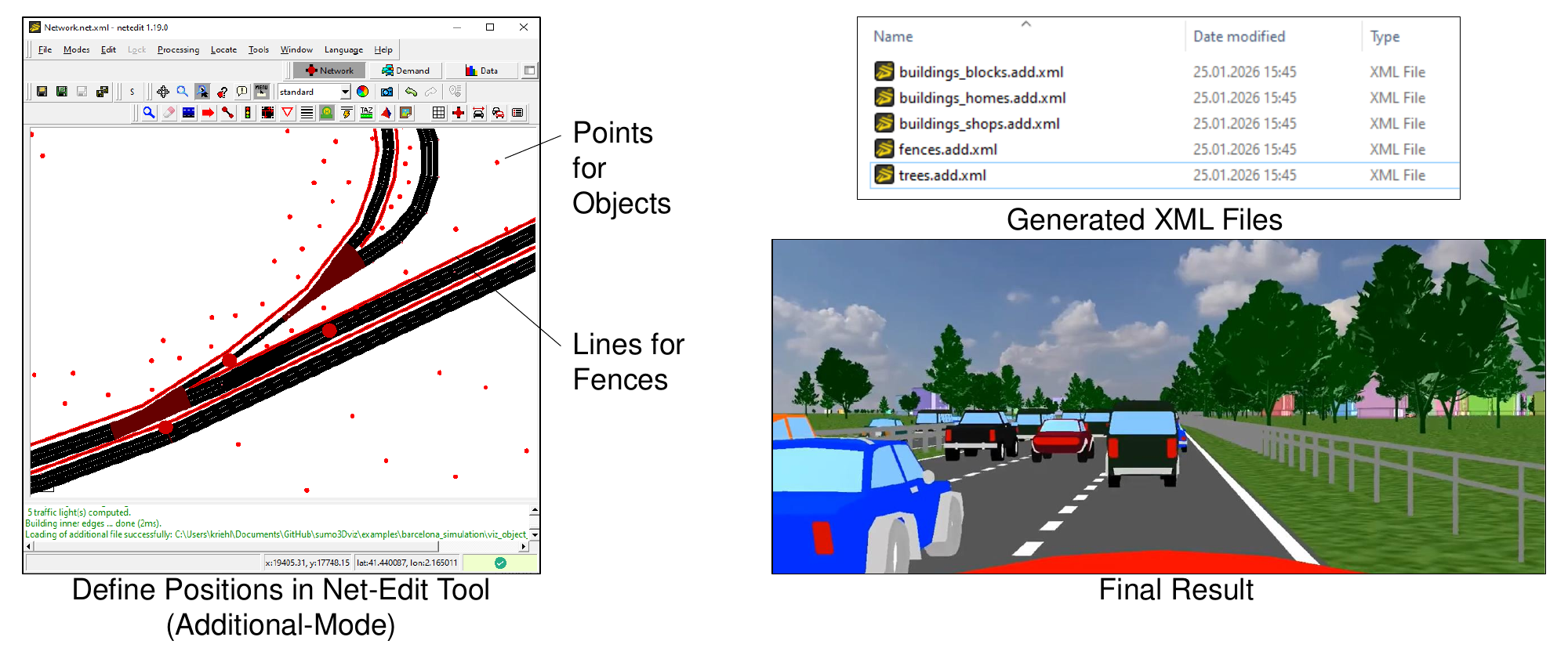}
    \caption{Positioning static objects for \texttt{sumo3Dviz} in \texttt{netedit}.}
    \label{fig:object_position}
\end{figure}

Static scene elements (trees, freeway fences, traffic lights, shops, homes, and building blocks) can be placed in the simulation, a selection of available models is shown in Figure~\ref{fig:static_objects}.
Their positions can be specified in SUMO polygon (POI) files, which can be edited using \texttt{netedit}, as shown in Figure~\ref{fig:object_position}. 

\FloatBarrier

\begin{figure}[!ht]
    \centering
    \includegraphics[width=1.0\linewidth]{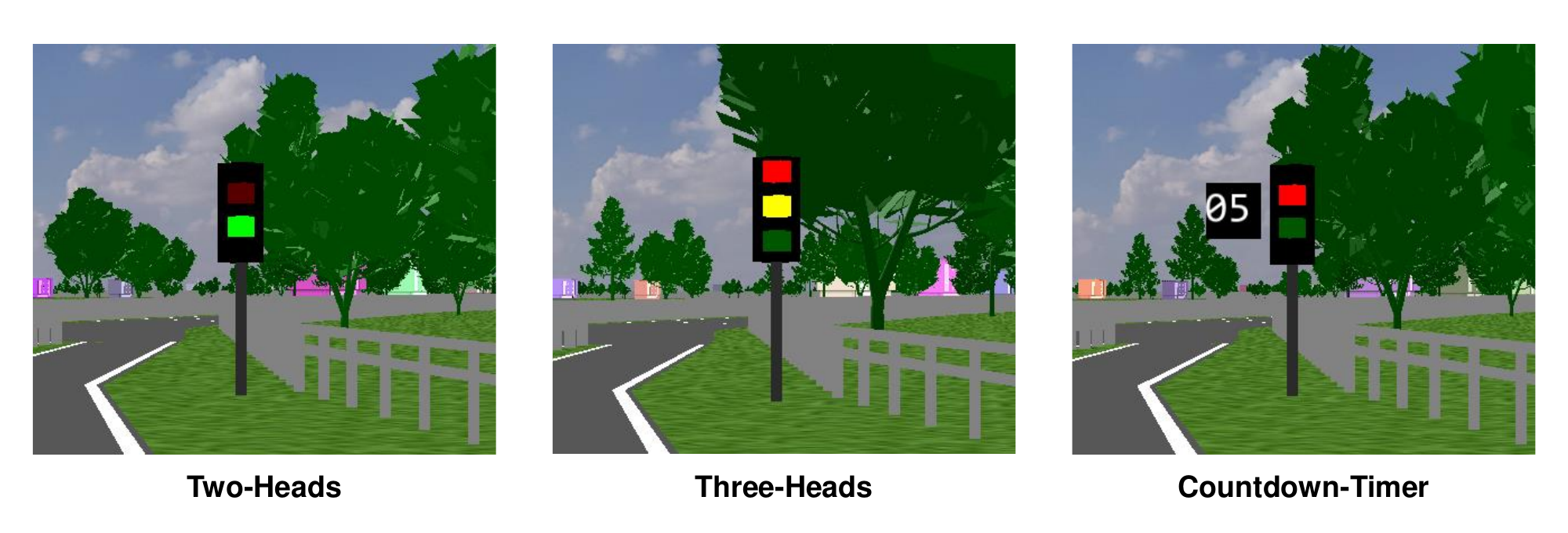}
    \caption{Traffic light designs in \texttt{sumo3Dviz}.}
    \label{fig:trafficlights}
\end{figure}

Traffic lights can be placed in the 3D visualisation as well.
Given the log files from the SUMO simulation, they can display the signal states at given times.
\texttt{sumo3Dviz} offers three traffic light designs, as shown in Figure~\ref{fig:trafficlights}, including two-heads (red and green), three-heads (red, yellow, and green, where yellow is interpolated automatically before and after a transition from green to red), and countdown timer.

\begin{figure}[!ht]
    \centering
    \includegraphics[width=1.0\linewidth]{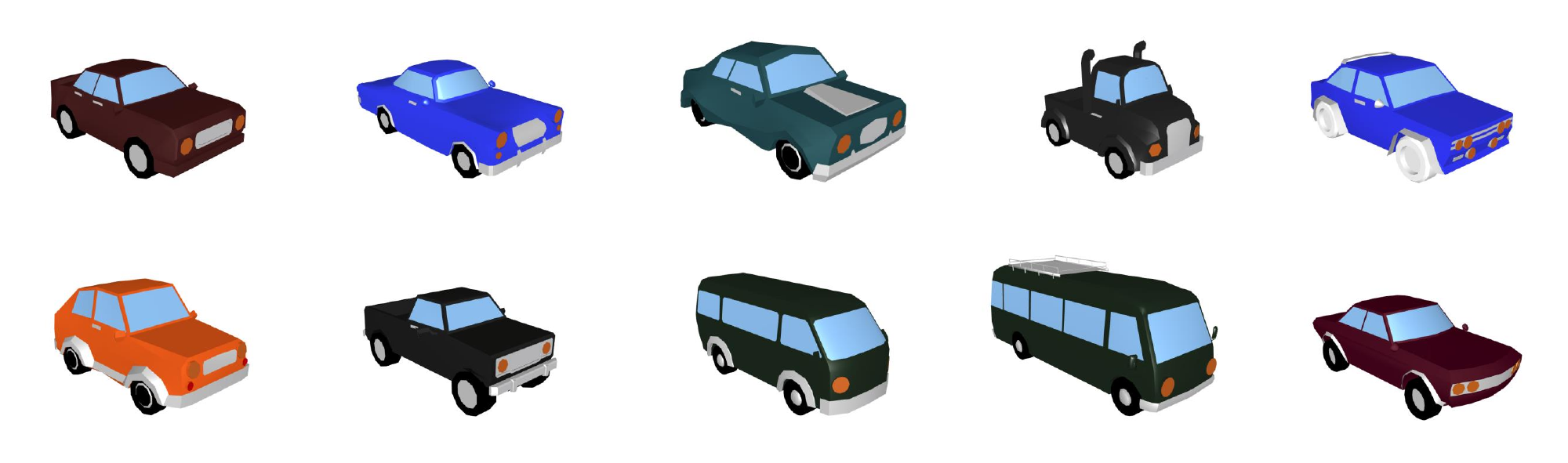}
    \caption{Car Models in \texttt{sumo3Dviz}.}
    \label{fig:car_models}
\end{figure}

In terms of moving objects (vehicles) a selection of ten car models are available, as shown in Figure~\ref{fig:car_models}.
Each vehicle reported in SUMO's trajectory log file is randomly assigned one of the car models, and then rendered in \texttt{sumo3Dviz}.

\FloatBarrier

\subsection{Trajectory Smoothing}

The Cartesian trajectory of a vehicle $i$ at discrete simulation time steps 
$k \in \{0,1,\dots,K\}$ is given by
\begin{equation}
    \mathbf{X}_k^i = 
    \begin{pmatrix}
        x_k^i \\
        y_k^i \\
        \alpha_k^i
    \end{pmatrix},
\end{equation}
where $(x_k^i, y_k^i)$ denotes the 2D position and $\alpha_k^i \in [0^\circ,360^\circ)$ the orientation angle.

The simulation provides trajectories at a low temporal resolution 
$\Delta t_k$ (e.g.\ $1\,\mathrm{Hz}$ in SUMO). 
For visualisation and video rendering, a temporally higher resolved trajectory 
$\mathbf{X}_t^i$ with time step $\Delta t = 1/\mathrm{FPS}$ 
(e.g.\ $25\,\mathrm{Hz}$) is required.

\paragraph{Step 1: Angle Unwrapping}

To avoid discontinuities at the $0^\circ/360^\circ$ boundary, 
the orientation signal is first unwrapped:
\begin{equation}
    \tilde{\alpha}_k^i = \mathrm{unwrap}(\alpha_k^i),
\end{equation}
where $\mathrm{unwrap}(\cdot)$ removes artificial $2\pi$ jumps in the angular representation.

\paragraph{Step 2: Linear Interpolation}

Let $t$ denote the high-resolution time index. 
For each $t$ between $k$ and $k+1$, linear interpolation is applied component-wise:

\begin{equation}
    \mathbf{Y}_t^i
    =
    \mathbf{X}_k^i
    +
    \frac{t - t_k}{t_{k+1} - t_k}
    \left(
        \mathbf{X}_{k+1}^i - \mathbf{X}_k^i
    \right) =
    \begin{pmatrix}
        x_t^i \\
        y_t^i \\
        \tilde{\alpha}_t^i
    \end{pmatrix}.
\end{equation}

\paragraph{Step 3: Rolling-Window Smoothing}

To let the trajectory appear more natural and less linear, a centred moving average with window size $L$ is applied:

\begin{equation}
    \mathbf{Z}_t^i
    =
    \frac{1}{|W_t|}
    \sum_{j \in W_t}
    \mathbf{Y}_j^i,
\end{equation}

where 
\[
W_t = \left\{ j \; \middle| \; |j - t| \le \frac{L-1}{2} \right\}
\]
denotes the centred window and $|W_t|$ its cardinality 
(reduced near the trajectory boundaries).
Depending on $L$, the level of smoothing can be controlled; here we found $L=21$ most suitable for a frame rate of $25\;Hz$.

\paragraph{Step 4: Velocity-Based Orientation Estimation}

While linear interpolation and rolling-window smoothing works well for the positional components of vehicles' trajectories, the orientational component is enhanced further in these steps, for more natural movements.
The orientation of the vehicle is especially not derived as a pure tangential to the positional trajectory.

The positional increments (velocities) $\Delta x_t^i$ and $\Delta y_t^i$ are computed as follows:

\begin{equation}
    \Delta x_t^i = x_t^i - x_{t-1}^i,
    \qquad
    \Delta y_t^i = y_t^i - y_{t-1}^i,
\end{equation}

with step distance

\begin{equation}
    d_t^i = \sqrt{(\Delta x_t^i)^2 + (\Delta y_t^i)^2}.
\end{equation}

For sufficiently large motion ($d_t^i > d_{\min}$), 
the movement direction is estimated by

\begin{equation}
    \hat{\alpha}_t^i
    =
    90^\circ
    -
    \mathrm{atan2}(\Delta y_t^i, \Delta x_t^i).
\end{equation}

If $d_t^i \le d_{\min}$, the orientation is retained from the smoothed interpolated signal to avoid numerical instability.

Finally, the estimated orientation is smoothed again using a centred moving average of window size $L_\alpha$:

\begin{equation}
    \alpha_t^i
    =
    \frac{1}{|W_t^\alpha|}
    \sum_{j \in W_t^\alpha}
    \hat{\alpha}_j^i.
\end{equation}

Here, we found the values $d_{min}=0.03$ and $L_\alpha = 41 ^\circ$ most suitable for achieving more realistic orientational movements.

\paragraph{Step 5: Final, Smoothed Trajectory}

The final high-resolution smoothed trajectory is therefore

\begin{equation}
    \mathbf{X}_t^i =
    \begin{pmatrix}
        x_t^i \\
        y_t^i \\
        \alpha_t^i
    \end{pmatrix}.
\end{equation}

This procedure ensures temporally smooth, visually consistent vehicle motion while preserving physically meaningful and visually appealing vehicle orientation dynamics.

\section{Case Study}

The capabilities of \texttt{sumo3Dviz} are demonstrated through a microsimulation-based case study of the \textit{Ronda de Dalt} metropolitan highway in Barcelona, Spain.
The example is designed to illustrate typical visualisation tasks rather than to evaluate traffic control performance.
The rendering process takes between $100 - 500~\mathrm{s}$ per frame on a conventional state-of-the-art laptop, after an initial loading procedure lasting for around $20~\mathrm{s}$.


\subsection{Context and Relevance}

The \textit{Ronda de Dalt} is one of Barcelona’s most critical road infrastructure elements.
Unlike conventional intercity motorways, it serves a multifunctional role by accommodating urban, metropolitan, and regional traffic within a dense urban environment~\cite{diaz2015espacio,riehl2025eq}.
Together with the \textit{Ronda Litoral}, it forms the primary ring road system known as \textit{Les Rondes}.
This system was constructed to redistribute traffic flows away from the urban core and improve citywide mobility.
Traffic volumes along this corridor have increased substantially in recent years.
At present, the \textit{Ronda de Dalt} operates without active ramp metering.
The case study~\cite{riehl2025eq} therefore explores a hypothetical ramp metering configuration to illustrate how such scenarios can be visualised and communicated using \texttt{sumo3Dviz}.


\begin{figure}[!h]
    \centering
    \caption{Network geometry of the \textit{Ronda de Dalt} highway network used in the case study, showing on-ramps, off-ramps, and residential areas.}
    \label{fig:map}
    \includegraphics[width=1.0\linewidth]{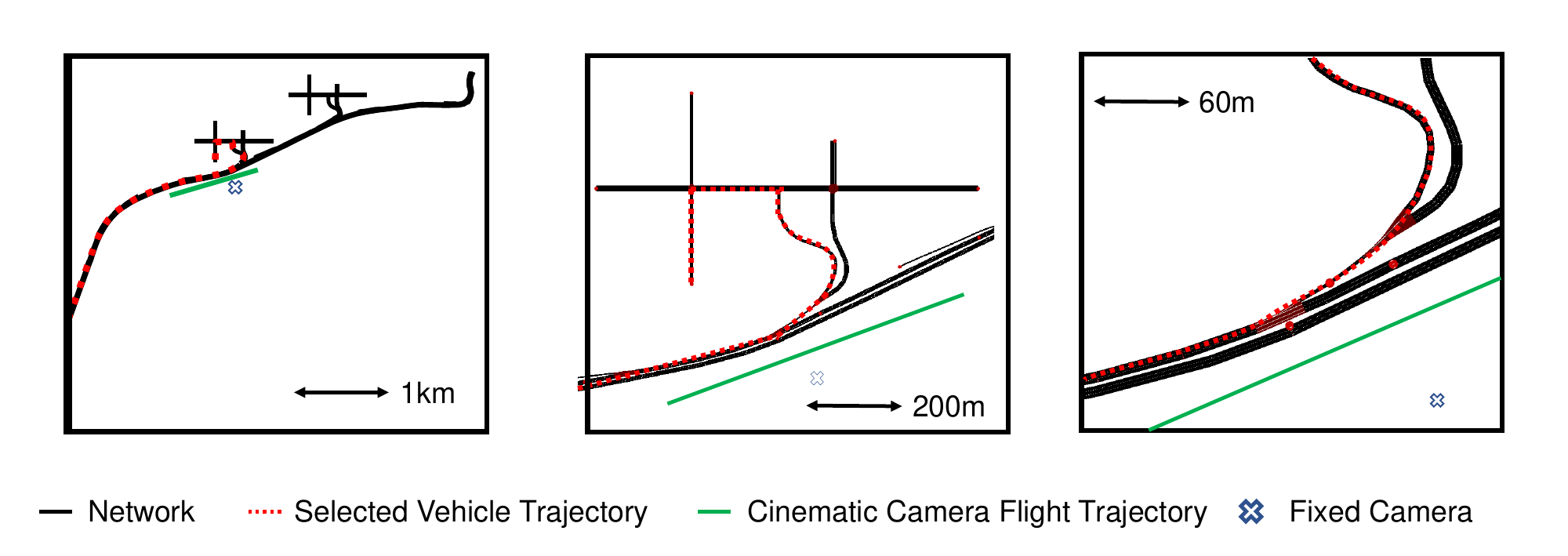}
\end{figure}
    
\subsection{Simulation Scenario}

The case study is based on a demand-calibrated SUMO simulation of the \textit{Ronda de Dalt} ring road~\cite{riehl2025eq}.
The simulated network spans approximately $6.5\,\mathrm{km}$ and consists of three lanes per direction with five on- and off-ramps, as well as two fictional residential areas around two on-ramps.
The simulation covers one hour of morning peak traffic between 06:00 and 07:00.
Traffic demand and calibration parameters are derived from Barcelona’s Open Data portal BCN~\cite{BarcelonaOpenData}.
To reproduce characteristic congestion patterns, vehicle interactions are modelled using the \textit{SL2015} lane-changing model~\cite{erdmann2015sumo}.
The vehicle population is divided into ten equally sized groups with varying levels of cooperation and aggressiveness.
All on-ramps are equipped with fixed-cycle ramp metering signals with a cycle length of $60\,\mathrm{s}$, consisting of $45\,\mathrm{s}$ green and $15\,\mathrm{s}$ red.


\subsection{Visualisation Setup}

For visualisation purposes, the SUMO simulation is extended with additional polygon and point-of-interest definitions (POIs).
These elements specify the locations of buildings, trees, and fences, providing minimal environmental context for three-dimensional rendering.
Vehicle position logging and signal state logging are enabled using standard SUMO logging and output options.
The network file, POIs, and simulation logs are consumed directly by \texttt{sumo3Dviz}; no modification of the simulation logic or post-processing of the output log files is required.


\subsection{Rendering Modes}

\begin{figure}[!h]
    \centering
    \caption{Visualisation modes of \texttt{sumo3Dviz}.}
    \label{fig:mode_viz}
    \includegraphics[width=1.0\linewidth]{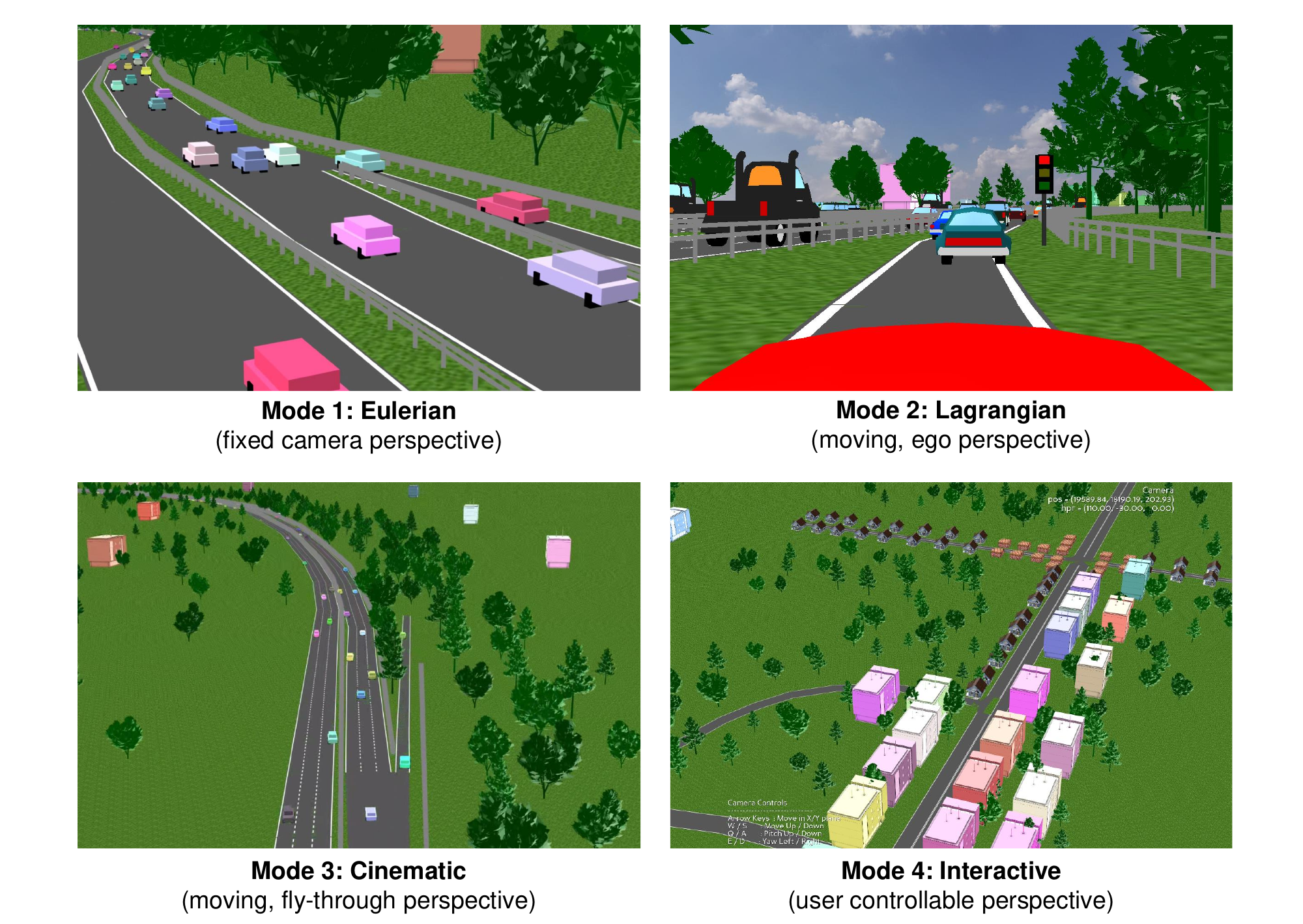}
\end{figure}

The case study demonstrates all four rendering modes supported by \texttt{sumo3Dviz}.
\begin{itemize}
    \item In the first mode, a fixed external camera is placed at a predefined location, providing an Eulerian overview of traffic conditions.
    \item In the second mode, the camera follows an individual vehicle along its trajectory, enabling an ego-centric perspective.
    \item The third mode renders a cinematic fly-through along a predefined three-dimensional camera path aligned with the highway.
    \item The fourth mode allows interactive navigation of the static scene using keyboard controls for camera position and orientation.
\end{itemize}
Each mode is illustrated using representative frames and video excerpts.
Figure~\ref{fig:mode_viz} presents example renderings for each visualisation mode.

\subsection{Visual Customisation}

The visualisations in this case study are rendered as video sequences.
Key rendering parameters include the simulated time interval, frame rate, image resolution, and camera configuration.
These parameters are specified through configuration files and can be adjusted without modifying the simulation or rendering code. When making use of the module-based components in a custom script, the options can be passed as function parameters.
The case study intentionally focuses on a limited set of options to illustrate typical usage.



\section{Conclusion}
This paper introduced \texttt{sumo3Dviz}, a lightweight and open-source three-dimensional visualisation pipeline for SUMO traffic simulations.
The tool is designed to bridge the gap between abstract two-dimensional visualisation and resource-intensive game-engine-based rendering approaches.
By processing standard SUMO simulation outputs directly, \texttt{sumo3Dviz} enables the reproducible and scriptable generation of three-dimensional traffic scenes accessible to users without extensive programming expertise.

The proposed approach prioritises simplicity, platform-independence, and integration into existing simulation workflows in Windows, macOS, and Linux distributions.
Support for multiple camera perspectives, batch rendering, and configurable visual styles enables applications in teaching, communication, and human-centred traffic research.
Trajectory interpolation and orientation smoothing ensure visually coherent motion from discrete simulation outputs, which is particularly important for perceptual evaluation.

A case study of the \textit{Ronda de Dalt} metropolitan highway in Barcelona demonstrated \texttt{sumo3Dviz}'s application to realistic, demand-calibrated microsimulations.
The example illustrated multiple rendering modes and highlighted the flexibility of the tool for different visualisation tasks.


\section*{Data availability statement}
The materials (textures, 3D models), source code, examples, and documentation can be found on the project's GitHub page \url{https://github.com/DerKevinRiehl/sumo3Dviz/} and the pip/PyPi archive page \url{https://pypi.org/project/sumo3Dviz/}.


\section*{Author contributions}
\begin{itemize}
    \item \textbf{Kevin Riehl:} Conceptualisation, Methodology, Software, Validation, Investigation, Resources, Writing - Original Draft, Visualisation, Project administration, Funding acquisition.
    \item \textbf{Julius Schlapbach:} Conceptualisation, Methodology, Software, Validation, Investigation, Resources, Writing - Review \& Editing, Visualisation.
    \item \textbf{Anastasios Kouvelas:} Writing - Review \& Editing, Supervision.
    \item \textbf{Michail A. Makridis:} Writing - Review \& Editing, Supervision, Funding acquisition.
\end{itemize}

\section*{Competing interests}
The authors declare that they have no competing interests.

\section*{Funding}
This work has received funding from the European Union’s Horizon Europe research and innovation programme under grant agreement No. 101203465 (project "FEDORA"), and from the Swiss State Secretariat for Education, Research and Innovation (SERI) under contract No. REF-1131-52301.

\newpage
\printbibliography

\end{document}